\documentclass[10 pt]{proposal_mine}
\usepackage{caption, float, graphics, graphicx, my_macros}

\title{A Hubble Astrometry Initiative: Laying the Foundation for the Next-Generation Proper-Motion Survey of the Local Group}

\begin{document}

\begin{center}
\textbf{\Large A Hubble Astrometry Initiative: Laying the Foundation for the Next-Generation Proper-Motion Survey of the Local Group}

\vspace{5 mm}

{\large White Paper for Hubble's 2020 Vision

\vspace{10 mm}

Nitya Kallivayalil (University of Virginia) \\
Andrew R.~Wetzel (Caltech \& Carnegie Observatories) \\
Joshua D.~Simon (Carnegie Observatories) \\
Michael Boylan-Kolchin (University of Maryland) \\
Alis J.~Deason (UC Santa Cruz) \\
Tobias K.~Fritz (University of Virginia) \\
Marla Geha (Yale University) \\
Sangmo Tony Sohn (Johns Hopkins University) \\
Erik J.~Tollerud (Yale University) \\
Daniel R.~Weisz (University of Washington) \\

\vspace{15 mm}

March 4, 2015
}

\end{center}

\vspace{15 mm}


\centerline{\textbf{\large Abstract}}

\vspace{5 mm}
\noindent
High-precision astrometry throughout the Local Group is a unique capability of the Hubble Space Telescope (HST), with potential for transformative science, including constraining the nature of dark matter, probing the epoch of reionization, and understanding key physics of galaxy evolution.
While Gaia will provide unparalleled astrometric precision for bright stars in the inner halo of the Milky Way, HST is the only current mission capable of measuring accurate proper motions for systems at greater distances ($\gtrsim 80 \kpc$), which represents the vast majority of galaxies in the Local Group.
The next generation of proper-motion measurements will require long time baselines, spanning many years to decades and possibly multiple telescopes, combining HST with the James Webb Space Telescope (JWST) or the Wide-Field Infrared Survey Telescope (WFIRST).
However, the current HST allocation process is not conducive to such multi-cycle/multi-mission science, which will bear fruit primarily over many years.
We propose an HST astrometry initiative to enable long-time-baseline, multi-mission science, which we suggest could be used to provide comprehensive kinematic measurements of all dwarf galaxies and high surface-density stellar streams in the Local Group with HST's Advanced Camera for Surveys (ACS) or Wide Field Camera 3 (WFC3).
Such an initiative not only would produce forefront scientific results within the next 5 years of HST's life, but also would serve as a critical anchor point for future missions to obtain unprecedented astrometric accuracy, ensuring that HST leaves a unique and lasting legacy for decades to come.

\newpage

\noindent
\textbf{Introduction.}
The field of astrometry is likely to produce fundamental advances in the coming decades.
The remarkable stability and resolution of the Advanced Camera for Surveys (ACS) and Wide Field Camera 3 (WFC3) on the Hubble Space Telescope (HST) position it to play a leading role in this work, enabling precise measurements of proper motions (PMs) that complete the full 6-dimensional orbital phase space of nearby objects.
HST astrometry already has led to breakthroughs such as (1) determining the orbit of the Large and Small Magellanic Clouds (LMC and SMC), (2) detecting internal rotation in the LMC, (3) measuring the tangential motion of M31, (4) dynamically measuring the mass of the Milky-Way (MW) halo, and (5) constraining the presence of intermediate-mass black holes in globular clusters.
Simultaneously, the Local Group (LG), with its dwarf galaxies and stellar streams, has emerged as a key testbed for $\Lambda$CDM, the epoch of reionization, and the physics of galaxy formation and evolution.
However, progress in all of these areas is limited significantly by the lack of comprehensive and robust PM measurements throughout the LG.
We propose an astrometric initiative for long-time-baseline, multi-mission science with HST over the next 5 years, to measure PMs or establish PM baselines for all dwarf galaxies and selected stellar streams out to $1 \mpc$.
Combined with rich existing HST data on these systems, this initiative will answer fundamental questions about galaxy evolution, reionization, and the nature of dark matter.

\vspace{4 mm}
\noindent
\textbf{Five science drivers that motivate a comprehensive proper-motion survey of the Local Group:}

\vspace{1.5 mm}
(1) \textit{Direct dynamical measurements of the mass of the Milky Way and M31.}
HST-based orbital motions of both satellites and streams provide \emph{direct} dynamical measurements of the mass profiles of the MW and M31.
However, current constraints come from either using only line-of-sight velocities \citep[e.g.,][]{Tollerud2012}, or using just one satellite with a very accurate PM measurement \citep{BoylanKolchin2013}, both of which are limited by systematics.
By contrast, PM measurements of \emph{all} known satellites/streams in the LG will probe the potentials of the MW and M31 as a function of radius and provide significantly tighter constraints and cross checks on mass modeling, tripling the dimensionality of measured velocity space relative to line-of-sight studies.

\vspace{1.5 mm}
(2) \textit{Understanding the physics of environment on satellite galaxies.}
The dwarf galaxies in the LG show a strikingly sharp and nearly complete transition within the virial radii ($300 \kpc$) of the MW and M31, towards elliptical/spheroidal morphology, little-to-no cold gas, and quenched star formation \citep{Einasto1974}.
Thus, the MW and M31 halos exert the \emph{strongest} environmental influence on their galaxy populations of any observed systems.
Promisingly, as Figure~1 shows, many dwarfs now have measured star-formation histories (SFHs) \citep{Weisz2014a, Brown2014}, based on a total of $\sim 1000$ orbits of HST data.
These SFHs represent a level of detail not possible for more distant galaxies.
However, to fully leverage these SFHs in context, we must know the orbital history of each satellite, and proper motions will provide this, again in a way not possible beyond the LG.
Thus, the \emph{unique} combination of both PM measurements and existing SFHs make the LG the most compelling laboratory in which to study environmental effects on galaxies.

\vspace{1.5 mm}
(3) \textit{Dwarf galaxies as probes of cosmic reionization.}
Ultra-faint galaxies provide promising and direct probes of the effects of reionization (at $z > 6$) on the evolution of dwarf galaxies, using the above recently measured SFHs.
However, a significant challenge to using ultra-faints as probes of reionization is knowing where they were at $z > 6$, to disentangle the effects of reionization from the environmental effects of the MW halo.
PM-based orbits are the \emph{only} way to constrain the locations of ultra-faints during reionization, revealing whether they were isolated or within the MW halo at that time \citep[e.g.,][]{Wetzel2015}.

\vspace{1.5 mm}
(4) \textit{Physical associations of dwarf galaxies and stellar streams.}
Significant debate persists regarding whether several dwarfs and/or streams are part of the same physical associations/sub-groups \citep[e.g.,][]{Watkins2013}.
Such ``satellites of satellites'' would have important implications for dwarf evolution, such as tidal stripping, merging, and the formation of dark-matter cores.
Furthermore, the lack of new MW satellites identified in the Pan-STARRS survey has strengthened the apparent polar configuration of satellites around the MW.
This result, combined with the discovery of an apparently thin disk of satellites around M31 \citep{Ibata2013}, challenges whether the spatial distribution and kinematics of dwarfs in the LG are consistent with $\Lambda$CDM predictions.
In both cases, PM measurements would definitively determine whether these ``planes of satellites'' are physical structures or merely chance alignments.

\vspace{1.5 mm}
(5) \textit{Internal kinematics of dwarf galaxies.}
The inner mass profile of dwarfs has emerged as perhaps the most important test of the nature of dark matter, as well as the strength of galactic feedback (e.g., the cusp/core and ``Too Big to Fail'' problems).
Unfortunately, constraining whether dwarf halos are cuspy using projected light profiles and radial velocities has proven difficult \citep[e.g.,][]{BreddelsHelmi2013}.
However, PMs of individual stars within dwarfs would break projection/orbital degeneracies \citep{Evslin2015} and thus enable measurements of their mass profiles \citep{Strigari2007}.
As Figure~2 shows, such internal kinematics already have been measured for the LMC \citep{vanderMarelKallivayalil2014}.
With advances in analysis techniques and longer baselines, HST alone has the potential to measure internal kinematics of dwarfs of the MW within 5 years, and combining with future missions will extend the baseline an additional $10+$ years.
This will guarantee that the 3-D motions of individual stars are measured with sufficient accuracy to determine the mass distribution in these most dark-matter dominated galaxies, but only if we establish an HST baseline now. This may be the \emph{only} feasible way to probe the nature of dark matter on sub-kpc scales.

\begin{figure}
\centering
\includegraphics[width = 0.99 \textwidth]{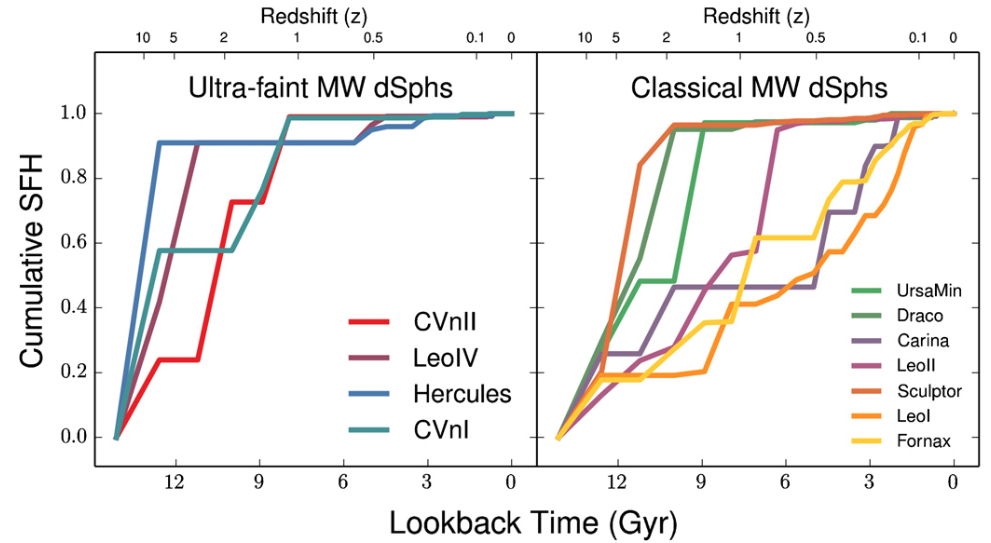}
\vspace{-2 mm}
\caption{
Cumulative star-formation histories (SFHs) for both ultra-faint and classical dwarf galaxies of the MW, selected from 40 galaxies with SFHs from \citet{Weisz2014a}, based on $\sim 1000$ orbits of HST imaging.
The ultra-faints are likely fossils of reionization, given that essentially all stars formed at $z > 3$ (or earlier), but the impact of the MW environment at that time remains unclear.
Proper motions directly will constrain an individual satellite’s orbital history and thus its location during and after reionization, to understand and disentangle the effects of reionization versus the MW halo environment on these SFHs.
}
\end{figure}

\begin{figure}
\centering
\includegraphics[width = 0.99 \textwidth]{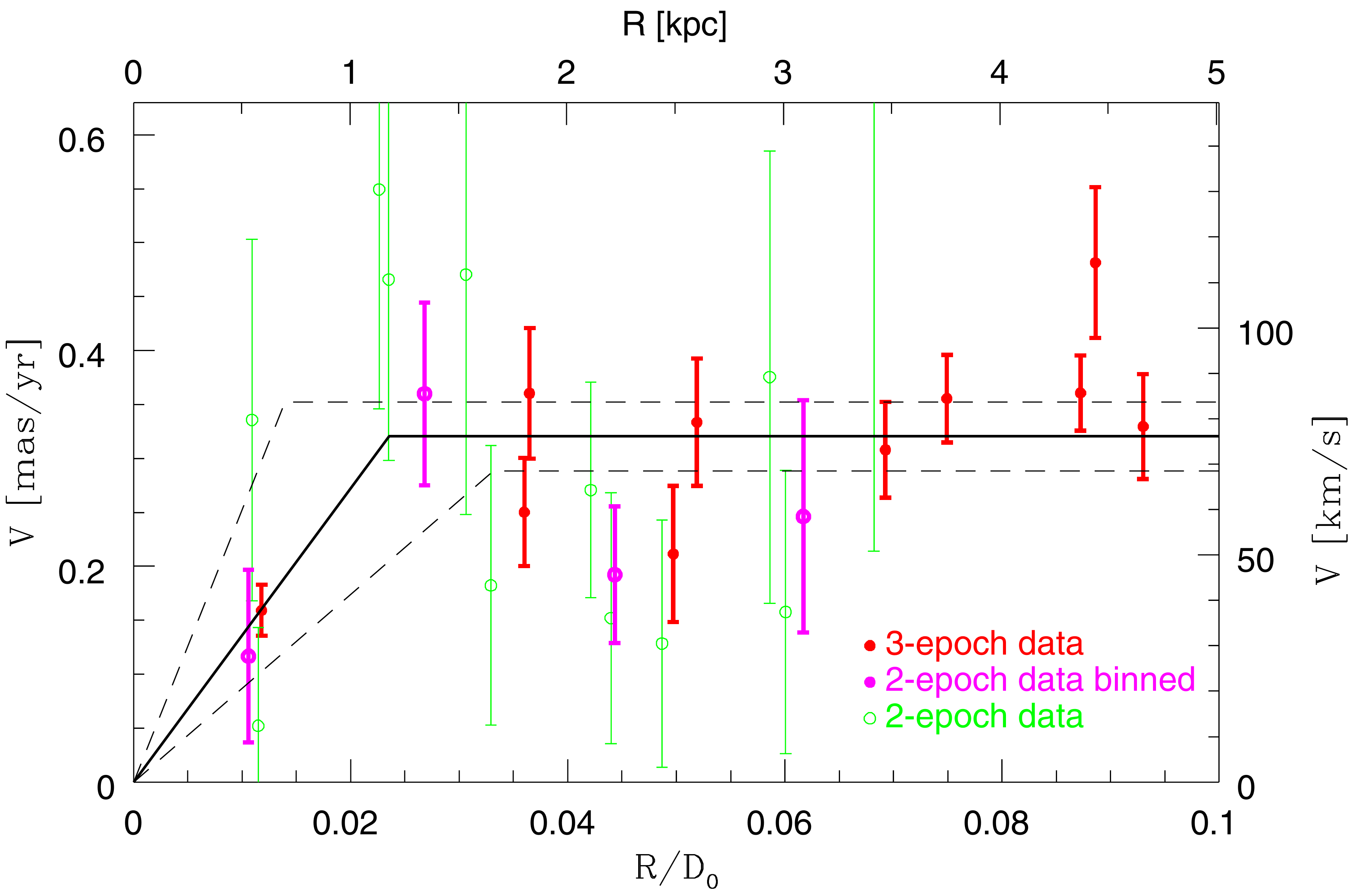}
\vspace{-2 mm}
\caption{
Internal kinematics of the LMC, from \citet{vanderMarelKallivayalil2014}.
Rotation velocity, inferred from proper motions with a 2-year (green) and 7-year (red) baseline, versus radius.
Left (right) axis shows angular (physical) velocity.
}
\end{figure}

\vspace{4 mm}
\noindent
\textbf{What about proper motions from Gaia?}
While Gaia is poised to provide unparalleled astrometric accuracy in the inner halo of the MW for bright stars ($V \lesssim 20$), Gaia is unlikely to provide sufficiently accurate PMs of individual satellites/streams beyond $\sim 80 \kpc$ to address the above science, leaving PM measurements at larger distances the sole province of larger-aperture facilities such as HST, the James Webb Space Telescope (JWST), and the Wide-Field Infrared Survey Telescope (WFIRST).
For example, for the Hercules dwarf (at $\sim 130 \kpc$), an optimistic PM error estimate for Gaia (which does not include systematic errors or correlated errors for neighboring stars) is $12 \kms$, insufficient to measure its internal dynamics.
With a 5-year baseline, HST not only can obtain better bulk PM errors ($\sim 9 \kms$) but also provides the \emph{only} avenue to measure internal PMs if combined with future missions.
Finally, even for some nearby ($\lesssim 80 \kpc$) dwarfs, their limited number of bright stars limits the precision with which Gaia can measure their PMs, so sufficiently deep HST measurements with long time baselines still provide the best PM accuracy.
In general, HST exceeds Gaia's performance for these systems for baselines longer than 3 years.

\vspace{4 mm}
\noindent
\textbf{Requirements for proper-motion measurements.}
A PM measurement requires at least two epochs, and the most important factors are (1) time baseline and (2) quality of \emph{each} epoch.
Only HST, using ACS or WFC3, has demonstrated the PM accuracy ($5 - 30 \kms$) needed for the above science \citep[e.g.,][]{Watkins2010}.
Based on existing work \citep{Sohn2013, Kallivayalil2013}, we expect an accuracy of 0.03 mas\,yr$^{-1}$ for a 5-year baseline, which translates to $4\,(40) \kms$ error at $30\,(300) \kpc$.
This is sufficient to address points (1) through (4) above within the MW halo.
For stellar streams, which are sparser and benefit less from $\sqrt{N_{\rm star}}$, we expect accuracies of 0.1 mas\,yr$^{-1}$ per pointing ($12 \kms$ at $25 \kpc$) \citep{Sohn2014}, though multiple pointings along a stream can strongly constrain the MW potential \citep[e.g.,][]{Kupper2015}.
Measuring either \emph{internal} kinematics of MW dwarfs to better than $5 \kms$ per star, or orbital PMs of dwarfs beyond $300 \kpc$ to within $30 \kms$ per dwarf, will require 10 to 20 years (depending on distance).
Thus, much of the key science will become possible only in the 2020's with HST (if still functional), JWST, and WFIRST (HST's resolution and depth are well matched to these missions), \emph{but only if HST establishes a firm baseline now}.
However, of the 70 known dwarfs out to $1 \mpc$, \emph{only} 10 have published PM measurements, and at least 18 lack even a reliable first-epoch baseline with ACS/WFC3. It is imperative that HST observe these systems as soon as possible.

\vspace{4 mm}
\noindent
\textbf{Estimated size of a proper-motion survey of the Local Group.}
Within $1 \mpc$, there are 70 dwarfs, and we consider here a subset of 6 (of the many) stellar streams/structures that have sufficiently high surface densities (Sgr, Pyxis, GD-1, Orphan, Pal 5, TriAnd).
Considering 5 pointings along each stream, this amounts to 100 targets in total.
To survey the LG, we suggest a two-phase approach: in Year 1, obtain one set of observations for all objects that do not already have sufficient ACS/WFC3 imaging (18 dwarfs and 2 of the above streams). In Year 5, obtain a second epoch for all targets that have sufficient archival ACS/WFC3 imaging, as well as for the 10 targets from Year 1 that are within $300 \kpc$. 
Previous work suggests that an average of 4 visits per target, depending on distance, would allow empirical self-calibration of systematic errors.
Thus, such a program would amount to a total of 440 orbits (112 in Year 1 and 328 in Year 5).
This phased approach both will allow HST alone to address many of the goals 1 - 4 above using the maximum time baseline possible and will ensure optimal first epochs to measure internal kinematics and orbits of M31 dwarfs with future missions.
Finally, we note that these numbers represent lower limits, because we expect ongoing deep photometric surveys, such as the Dark Energy Survey or Hyper Suprime-Cam Survey, to discover many new dwarfs and streams, so we also suggest an avenue for allocating time over the next 5 years to obtain first-epoch PM imaging for these.

\vspace{3 mm}
\noindent
\textbf{Limitations of the current HST allocation process.}
The current HST General Observer (GO) allocation process is not conducive to enabling such multi-year/multi-decade experiments.
In particular, no mechanism exists for obtaining multiple epochs separated by many years, or for projects whose primary science goals will require combining HST observations with data to be obtained by future telescopes, even if the scientific returns are profound and fundamental.
The current GO process is effective only for limited types of astrometric projects: (1) those that can be completed using two epochs within the 3-year timeline of a long-term GO program, which necessarily restricts PM measurements to nearby systems and to \emph{bulk} proper motions rather than internal kinematics, and (2) those that can be completed by obtaining a second epoch for a field that already has first-epoch data in the archive from $5+$ years ago, which restricts PM measurements mostly to the more luminous ``classical'' dwarfs.
As a result, current projects are limited to a handful of objects with limited time baselines and limited science drivers.

\vspace{4 mm}
\noindent
\textbf{Solution: an astrometry initiative for long-time-baseline and multi-mission projects.}
To address the above limitations, we propose a mechanism to give substantial weight not just to the immediate scientific return from HST projects, but also to the science that HST data will enable in the future, when combined with future data from HST or other facilities. Specifically, we propose that HST launch an astrometry initiative, analogous to the current UV initiative, to promote PM science. HST is uniquely suited for this application, but it necessarily spans multiple years/decades and multiple missions/telescopes. More broadly, we also propose that HST develop a separate allocation category (at least at the $\sim 1000$ orbit level over several cycles) specifically for long-time-baseline projects and/or those that will be realized together with data from future missions, such as JWST or WFIRST, in the years/decades to come. While such initiatives will help to enable groundbreaking PM science in the LG, they more generally will open a broad class of long-time-baseline science programs that can be extended for decades with future missions to obtain unprecedented accuracy, ensuring that HST leaves a unique and lasting legacy to the 2020's and beyond.


\vspace{8 mm}

\bibliography{white_paper}

\begin{thebibliography}{}
\expandafter\ifx\csname natexlab\endcsname\relax\def\natexlab#1{#1}\fi

\bibitem[{{Boylan-Kolchin} {et~al.}(2013){Boylan-Kolchin}, {Bullock}, {Sohn},
  {Besla}, \& {van der Marel}}]{BoylanKolchin2013}
{Boylan-Kolchin}, M., {Bullock}, J.~S., {Sohn}, S.~T., {Besla}, G., \& {van der
  Marel}, R.~P. 2013, \apj, 768, 140

\bibitem[{{Breddels} \& {Helmi}(2013)}]{BreddelsHelmi2013}
{Breddels}, M.~A., \& {Helmi}, A. 2013, \aap, 558, A35

\bibitem[{{Brown} {et~al.}(2014){Brown}, {Tumlinson}, {Geha}, {Simon},
  {Vargas}, {VandenBerg}, {Kirby}, {Kalirai}, {Avila}, {Gennaro}, {Ferguson},
  {Mu{\~n}oz}, {Guhathakurta}, \& {Renzini}}]{Brown2014}
{Brown}, T.~M., {Tumlinson}, J., {Geha}, M., {et~al.} 2014, \apj, 796, 91

\bibitem[{{Einasto} {et~al.}(1974){Einasto}, {Saar}, {Kaasik}, \&
  {Chernin}}]{Einasto1974}
{Einasto}, J., {Saar}, E., {Kaasik}, A., \& {Chernin}, A.~D. 1974, \nat, 252,
  111

\bibitem[{{Evslin}(2015)}]{Evslin2015}
{Evslin}, J. 2015, ArXiv e-prints, arXiv:1501.07503

\bibitem[{{Ibata} {et~al.}(2013){Ibata}, {Lewis}, {Conn}, {et~al.}}]{Ibata2013}
{Ibata}, R.~A., {Lewis}, G.~F., {Conn}, A.~R., {et~al.} 2013, \nat, 493, 62

\bibitem[{{Kallivayalil} {et~al.}(2013){Kallivayalil}, {van der Marel},
  {Besla}, {Anderson}, \& {Alcock}}]{Kallivayalil2013}
{Kallivayalil}, N., {van der Marel}, R.~P., {Besla}, G., {Anderson}, J., \&
  {Alcock}, C. 2013, \apj, 764, 161

\bibitem[{{K{\"u}pper} {et~al.}(2015){K{\"u}pper}, {Balbinot}, {Bonaca},
  {Johnston}, {Hogg}, {Kroupa}, \& {Santiago}}]{Kupper2015}
{K{\"u}pper}, A.~H.~W., {Balbinot}, E., {Bonaca}, A., {et~al.} 2015, ArXiv
  e-prints, arXiv:1502.02658

\bibitem[{{Sohn} {et~al.}(2013){Sohn}, {Besla}, {van der Marel},
  {Boylan-Kolchin}, {Majewski}, \& {Bullock}}]{Sohn2013}
{Sohn}, S.~T., {Besla}, G., {van der Marel}, R.~P., {et~al.} 2013, \apj, 768,
  139

\bibitem[{{Sohn} {et~al.}(2014){Sohn}, {van der Marel}, {Carlin}, {Majewski},
  {Kallivayalil}, {Law}, {Anderson}, \& {Siegel}}]{Sohn2014}
{Sohn}, S.~T., {van der Marel}, R.~P., {Carlin}, J.~L., {et~al.} 2014, ArXiv
  e-prints, arXiv:1408.3408

\bibitem[{{Strigari} {et~al.}(2007){Strigari}, {Bullock}, \&
  {Kaplinghat}}]{Strigari2007}
{Strigari}, L.~E., {Bullock}, J.~S., \& {Kaplinghat}, M. 2007, \apjl, 657, L1

\bibitem[{{Tollerud} {et~al.}(2012){Tollerud}, {Beaton}, {Geha}, {Bullock},
  {Guhathakurta}, {Kalirai}, {Majewski}, {Kirby}, {Gilbert}, {Yniguez},
  {Patterson}, {Ostheimer}, {Cooke}, {Dorman}, {Choudhury}, \&
  {Cooper}}]{Tollerud2012}
{Tollerud}, E.~J., {Beaton}, R.~L., {Geha}, M.~C., {et~al.} 2012, \apj, 752, 45

\bibitem[{{van der Marel} \&
  {Kallivayalil}(2014)}]{vanderMarelKallivayalil2014}
{van der Marel}, R.~P., \& {Kallivayalil}, N. 2014, \apj, 781, 121

\bibitem[{{Watkins} {et~al.}(2010){Watkins}, {Evans}, \& {An}}]{Watkins2010}
{Watkins}, L.~L., {Evans}, N.~W., \& {An}, J.~H. 2010, \mnras, 406, 264

\bibitem[{{Watkins} {et~al.}(2013){Watkins}, {van de Ven}, {den Brok}, \& {van
  den Bosch}}]{Watkins2013}
{Watkins}, L.~L., {van de Ven}, G., {den Brok}, M., \& {van den Bosch},
  R.~C.~E. 2013, \mnras, 436, 2598

\bibitem[{{Weisz} {et~al.}(2014){Weisz}, {Dolphin}, {Skillman}, {Holtzman},
  {Gilbert}, {Dalcanton}, \& {Williams}}]{Weisz2014a}
{Weisz}, D.~R., {Dolphin}, A.~E., {Skillman}, E.~D., {et~al.} 2014, \apj, 789,
  147

\bibitem[{{Wetzel} {et~al.}(2015){Wetzel}, {Deason}, \&
  {Garrison-Kimmel}}]{Wetzel2015}
{Wetzel}, A.~R., {Deason}, A.~J., \& {Garrison-Kimmel}, S. 2015,
  arXiv:1501.01972

\end{thebibliography}

\end{document}